\documentclass[pre,preprintnumbers,twocolumn,showpacs]{revtex4-1}
\usepackage{graphicx,t1enc}
\usepackage{amssymb,amsmath}
\usepackage{color}
\usepackage{units}
\usepackage[caption=false]{subfig}
\usepackage[titletoc]{appendix}
\begin{document}

\title{Statistical fluctuations in pedestrian evacuation times and the effect
of social contagion}

\author{Alexandre Nicolas}\email[]{alexandre.nicolas@polytechnique.edu}
\author{Sebastián Bouzat}\email[]{bouzat@cab.cnea.gov.ar}
\author{Marcelo N. Kuperman}\email[]{kuperman@cab.cnea.gov.ar}

\affiliation{Consejo Nacional de Investigaciones Cient\'{\i}ficas y T\'ecnicas \\
Centro At\'omico Bariloche (CNEA) and Instituto Balseiro, R8400AGP Bariloche, Argentina.}

\begin{abstract}
Mathematical models of pedestrian evacuation and the associated simulation
software have become essential tools for the assessment of the safety of public
facilities and buildings. While a variety of models are now available, their
calibration and test against empirical data are generally restricted to global,
averaged quantities;  the statistics compiled from the time series  of
individual escapes (``microscopic'' statistics) measured in recent experiments
are thus overlooked. In the same
spirit, much research has primarily focused on the average global evacuation
time, whereas the whole distribution of evacuation times over some set of
realizations should matter. In the present paper we propose and discuss the
validity of a simple relation between this distribution and the ``microscopic''
statistics, which is theoretically valid in the absence of correlations. To this
purpose, we develop a minimal cellular automaton, with novel features that
afford a
semi-quantitative reproduction of the experimental ``microscopic'' statistics.
We then introduce a process of social contagion of impatient behavior in the
model and show that the simple relation under test may dramatically fail at high
contagion strengths, the latter being responsible for the emergence of strong
correlations in the system. We conclude with comments on the potential practical
relevance for safety science of calculations based on ``microscopic''
statistics.
\end{abstract}


\date{\today}

\maketitle

\section{I. Introduction}

Large buildings or public facilities should allow a safe and quick evacuation of
the attendance in the event of an emergency, such as a fire. Indeed, the
uncontrolled movement of large crowds may involve excessive delays due to
obstruction at narrowings and, in the most extreme cases, presents a risk of
injury or even death because of pushing and trampling. To limit these risks, the
design and construction of public facilities must obey strict standards,
specified in building codes. In addition to
design criteria that can render environments safer, a reliable
prediction tool for the total evacuation time $T_{\mathrm{esc}}$, as a function,
e.g., of the number $N$ of
attendants, would be extremely valuable. To this end, a better understanding of
the dynamical processes governing crowd motion might be crucial.

In spite of being the primary concern of much research, the mean 
evacuation time $\bar{T}_{\mathrm{esc}}(N)$ does 
not convey enough information to assess the safety of a facility: the fact that 
$\bar{T}_{\mathrm{esc}}(N)$ lies in safe bounds does not tell us how often the evacuation will be 
excessively long. Fluctuations are indeed expected to be
large in such complex finite-size systems. Therefore one should consider the
whole distribution of total evacuation times 
over some (uncontrolled) space of realizations, i.e., for different compositions
of the 
attendance, states of mind, times of the day, etc.

An evacuation may be delayed by various possible factors, but a bottleneck that
turned fatal many times is the congested 
passage through an (insufficiently wide) exit door. Amidst many other similar tragedies, severe 
congestion at an exit or a narrowing was reported in the 1863 Church of the
Company
of Jesus fire catastrophe in Santiago (Chile) \cite{nytimes1864chile}, causing more than 2,000 
fatalities, in the 1903 Iroquois Theatre fire disaster in Chicago, with a death toll above 600 
\cite{pan2006human} (for the full story, see \cite{hatch2003tinder}), in the largely documented 
1989 crowd disaster at the Hillsborough stadium 
(England) \cite{nicholson1995investigation,dickie1995major} (in this case, congestion actually 
occurred at 
the entrance, under the pressure of the incoming crowd of supporters), in the 2010 crowd disaster 
during the Love Parade in Duisburg (Germany) \cite{helbing2012crowd}, where the same tunnel was 
used as both the entrance and the exit from the premises, in the 2004 fire disaster in the 
nightclub República de Cromañón in Buenos Aires \cite{wortman2005tragedia}, as well as in the 
recent fire disaster in the Colectiv nightclub in Bucharest in October 2015 
\cite{guardian2015bucharest}, where clear problems with the emergency exits were reported.

Recently, quantitative analogies have been brought to light by Zuriguel and
colleagues between the flow of 
grains through a small orifice and pedestrian evacuation through a narrow door, at least in the 
controlled conditions in which the experiments were conducted
\cite{Zuriguel2014clogging,Pastor2015experimental}, thus suggesting that the
process is dominated by 
basic physical mechanisms. The escape dynamics were probed ``microscopically'' (i.e., at the level 
of the individual); in particular, statistically, the distribution of time lapses $\Delta t$ 
between successive egresses was shown to be well described by a power law at long jam durations 
$\Delta t$, regardless of the behavior prescribed to the participants. If the bulk evacuation can 
indeed be robustly characterized microscopically, the following approach seems
very promising:

\begin{itemize}
 \item  compute the statistics of time lapses associated
with a \emph{given} exit geometry (``microscopic'' statistics), using existing
and future 
recordings of real emergency evacuations (or, for want of better, evacuation drills in conditions 
as realistic as ethically possible), and

 \item infer the sought-after distribution of $T_{\mathrm{esc}}(N)$
(``macroscopic'' distribution)
from the microscopic one, as a sum of random time lapses.

\end{itemize}

Here, we wonder about the validity and relevance of such tempting connection
between the microscopic 
statistics and the macroscopic distribution, henceforth called ``micro-macro relation'' for 
brevity. To explore this problem and illustrate our findings, we have developed a highly 
economical cellular automaton model that is loosely inspired from the granular analogy and, for the 
first time, reproduces semi-quantitatively the experimental data of \cite{Pastor2015experimental}. 
In particular, we show that the micro-macro relation may break down, notably owing to the 
empirically established possibility of psychosocial contagion (or
``behavioral 
imitation'', 
i.e., an enhanced tendency to behave competitively for neighbors of aggressive 
people).

In Section II we clarify the theoretical underpinning of the micro-macro connection and expose 
possible causes leading to its violation. Section III is dedicated to the presentation of the 
model, which is then tested against experimental data in Section IV in the absence of 
social contagion. In Section V, we implement the latter effect in the model in a
simple way and 
study its consequences. We conclude by critically commenting on the potential interest for safety 
science of the micro-macro relation, beyond our specific implementation.

\section{II. Theoretical exposure of the problem}

Suppose that a given facility needs to be evacuated. Schematically, the
individual escape towards a safe point may be decomposed into the following
phases
\begin{itemize}
\item[(a)]  a delay before reacting to the emergency (for various possible
reasons),
 \item[(b)] a phase of relatively unconstrained motion,
alone or in group,
\item[(c)] possible delays due to obstruction at narrowings and exits.
\end{itemize}
 
Here, we exclusively focus on the last point. To this day, the question of whether this is actually 
the bottleneck in tragic evacuations remains controversial. A significant
portion of the social 
psychology literature of the last decades questions the \emph{prevalence} of
competitive moves, that is, 
selfish rushes towards the exit causing clogs
\cite{johnson1987panic,mawson2005understanding,aguirre2005emergency,drury2009cooperation,
drury2009everyone}, and hints at pieces of evidence of conserved social norms
and 
cohesive behaviors, such as the will to assist fallen people, in
emergencies. Yet, reports on crowd 
tragedies (see references above) make the \emph{occurrence} of selfish rushes in
some situations
unquestionable. In fact,
even if, intrinsically, the individual pedestrians are willing to cooperate, we shall see with our 
simple model that a global competitive response can emerge and propagate in some cases.

Recently, the escape dynamics through a narrow ($\approx$70 cm-wide) door were
studied experimentally in controlled evacuation drills
\cite{Garcimartin2014experimental,Pastor2015experimental}. Three distinct
degrees of competitiveness were 
successively prescribed to the participants, from the prohibition of any contact
in the most cooperative settings to the permission of moderately soft pushes to elbow one's way, in 
the most competitive settings. In each case, for a fine characterization of the dynamics, the 
distribution $p(\Delta t)$ of time intervals $\Delta t$ between successive escapes was computed, 
with seemingly robust bulk statistics; similarly to the case of granular hopper flows, the 
distributions were found to be well described by power laws at large $\Delta t$
viz.,
\begin{equation}
p(\Delta t)\propto\Delta t^{-\alpha},\label{eq:power-law_lapses}
\end{equation}
where the exponent $\alpha$ decreases with increasing 
crowd competitiveness. Thus, long clogging events were more likely in the more
competitive crowds, which resulted in longer evacuation times. Note that only
values of $\alpha$ larger than 3 were
measured; this implies 
that $\Delta t$ has a well defined mean value and standard deviation. From this ``microscopic''
characterization of relatively competitive egresses, one may aspire to derive the practically 
relevant distribution $P_{N}\left(T_{\mathrm{esc}}\right)$ of global evacuation times of $N$ 
evacuees over an uncontrolled space of evacuation realizations, where
$T_{\mathrm{esc}}(N)=\sum_{i=1}^{N}\Delta t_{i}$ and 
$\Delta t_{i}$ is the time lapse between the $(i-1)^\mathrm{th}$ and
$i^\mathrm{th}$ escape (for
convenience, the $0^\mathrm{th}$ escape is defined as the start of the
evacuation). To
do so, one can think of 
$T_{\mathrm{esc}}$ as a sum of $N$ independent random time-lapse variables
$\Delta t$ and write
\begin{equation}
P_{N}\left(T_{\mathrm{esc}}\right)=p^{*N}\left(T_{\mathrm{esc}}\right),
\label{eq:random_sum}
\end{equation}
where the superscript $*N$ should be understood as a convolution product. 
\emph{Regardless of the nature of the distribution} $p(\Delta t)$, Eq.~\ref{eq:random_sum} yields a 
direct connection between the microscopic statistics and the global
distribution. In particular, if the mean value $\overline{\Delta t}$ and the
standard 
deviation $\sigma_t$ of $p(\Delta t)$ are well defined, the central limit
theorem implies 
that, in the limit of large attendance $N\gg1$,
\begin{equation}
P_{N}\left(T_{\mathrm{esc}}\right)\sim\mathcal{N}\left(N\overline{\Delta
t},\sqrt{N}\sigma_t\right),\label{eq:P_N_Gaussian}
\end{equation}
where $\mathcal{N}\left(m,\sigma\right)$ denotes the normal law of mean $m$ and standard deviation 
$\sigma$ \footnote{Some details about the limit distribution expected in the
hypothetical case of a power law with exponent $\alpha \leqslant 3$ are
provided in Appendix A.}.

Theoretically, it is well known that the micro-macro connection of Eq.~\ref{eq:random_sum} fails 
if the $\Delta t_{i}$ display strong 
correlations, instead of being independent. Yet, for practical purpose, it is
tempting to discard this mathematical caveat. Indeed, it is virtually
impossible to collect sufficient
data on $P_{N}$, whereas the microscopic statistics $p$ could
readily be compiled from 
recordings of real emergency evacuations in the studied geometry (or, all ethical issues left 
aside, competitive evacuation drills). In how far is this neglect
justified in practice?

\smallskip

Here, we claim that, beyond the variations in the crowd composition and the correlations in 
pressure in the crowd, a major limitation to the micro-macro relation originates in the propagation 
of non-cooperative, alarmed behaviors \footnote{This contagion process is often
referred to as panic-spreading, but the 
denomination may be improper \cite{aguirre2005emergency}}. Although this process
of behavioral 
contagion, whether deliberate or not, may be less widespread than traditionally thought (or even 
perhaps reported \cite{johnson1987panic}), it did occur in major crowd disasters and very generally 
aggravated the situation. Therefore, it should be heeded in the extreme conditions where safety is 
most imperiled. A recent example that demonstrates this effect is a video
showing
the beginning of a crowd 
movement at a gathering following the November, 13th 2015 terrorist attacks in
Paris \footnote{Video available at
https://www.youtube.com/watch?v=u2LOmnldQkk}; another arresting example is the
following 
statement by Marshall, based on his combat experience
during World War II: ``Every large panic starts with some very minor event... Troops will 
always run if they see others running and do not understand why''
\cite{marshall1947men}. 
Since competitive rushes towards the exit have an impact on the microscopic distribution $p$, 
contagion induces correlations that may strongly affect the global distribution $P_{N}$.

Let us illustrate this point with an extreme example. Suppose that the crowd is
extremely 
susceptible to fear; for instance, consider the two foregoing
examples of suggestible crowds. In 
the (unlikely) event of someone actually displaying signs of alarm, fear will
quickly pervade 
the whole crowd. As a result, the evacuation will be either slow, if someone
actually gets ``panicky'', or fast, in the opposite case. Accordingly, the
distribution of global evacuation times 
cannot be inferred from the sole microscopic distribution $p(\Delta t)$,
because the latter, as an average, 
mingles escape data for different crowd states. In particular, the micro-macro
relation would 
predict far rarer sluggish evacuations than actually occur.

\section{III. Presentation of the Contagion-Free Model}

To test these ideas in a more concrete framework, we wish to develop a minimalistic model for 
pedestrian evacuation dynamics, focusing on congestion effects (point (c)
above). The model should reproduce the following microscopic statistical
features suggested by experiments 
\cite{Zuriguel2014clogging}, without obfuscating the picture with complex modeling details:

\begin{itemize}
\item[(i)] a broader-than-exponential, power-law-like tail in the distribution
$p(\Delta t)$ for narrow doors, as in Eq.~\ref{eq:power_law_dist},
with exponents comparable to empirical values,
\item[(ii)] an exponential-like distribution of burst sizes, where a burst
is defined as a series of uninterrupted escapes.
\end{itemize}

Inspired by previous work 
\cite{Bouzat2014game}, we develop a lattice-based cellular automaton.
Importantly, its key ingredients can loosely be interpreted in
the context of granular flows, given the aforementioned analogies
\cite{Zuriguel2014clogging} (also see 
\cite{chen2013scaling}), but we account for the fact that unlike fluid
particles pedestrians are responsive to the environment and may take decisions
and
move accordingly. The semi-quantitative agreement attained by our model with
respect to the 
specific statistical features (i) and (ii) is, as far as we know, unprecedented in cellular 
automata (see Appendix~C for remarks pertaining to the
observation of power laws, notably in previous works).

In a nutshell, agents (representing pedestrians) will be positioned on the
cells of the lattice grid, with at most one agent per site, due to steric
constraints. At each time step, agents 
target, and may move to, one of the adjacent sites. The chosen direction is 
controlled by a static floor field that directs them towards the exit
\cite{Burstedde2001simulation}. We insist that the model is deliberately
minimalistic. Among other
simplifications, we make no attempt to describe the architecture of a real facility, or to account 
for the existence of social bonds in the evacuating crowd, even though they may be important in 
practice \cite{moussaid2010walking,mawson2005understanding}.

%\subsection{Geometry and floor field}
\smallskip

We opt for a simple geometry, namely, a square of side $L$, with a single door of width $L_{d}$ in 
the middle of one of the walls. Space is divided into square cells, of unit length. 

At each time step, each agent targets one of the (at most) four adjacent sites,
the so called von 
Neumann neighborhood, or chooses to stay at their current location. The
probability of selection of a site
 depends on its attractiveness, quantified by the absolute static floor field

\begin{equation}
A_{s}(x,y)=\underset{\text{proximity to 
target}}{\bar{d}-\underbrace{\sqrt{\left(x-x_{T}\right)^{2}+\left(y-y_{T}\right)^{2}}}},\label{
eq:static_floor_field}
\end{equation}
where $(x_{T},y_{T})=\left(\frac{L}{2},-L_{d}\right)$ are the coordinates of the targeted ``safe 
point'' behind the door and $\bar{d}$ is a large value that maintains the attractiveness positive 
in all cases. Conceptually, $A_{s}(x,y)$ can be regarded as the negative of a potential energy in a 
granular system. The floor field could naturally be refined by describing, e.g., wall 
effects, but our simple choice turned out to be satisfactory for the present study.

Furthermore, occupied sites should always be less attractive than free ones. Thus the static 
attractiveness is complemented with a dynamic part: should a site be currently occupied by 
another pedestrian, its attractiveness will be penalized by a large constant value 
$H_{0}=-10$, viz., $A=A_{s}-H_{o}$, so that free sites ($A=A_s$) will virtually
always
be preferred to occupied ones.

%\subsection{Selection of the target site}

A priori, pedestrians should always try to move to the adjacent free site of
highest attractiveness. However, this deterministic local rule traps the system
in metastable 
configurations, with unrealistic density profiles in front of the exit. Accordingly,
some stochasticity is introduced, in the form of a small amount $\eta$ of
noise
(which would be a 
measure of the amplitude of the vibrations in a granular system). A pedestrian,
on site $\mu$, 
selects a site $\nu$ in his/her vicinity (including the current site), with a
probability 
\begin{equation}
p_{\mu \rightarrow
\nu}\equiv\frac{e^{\frac{A_{\nu}-A_{\mu}}{\eta}}}
{\sum_{\sigma \in \Lambda_\mu
\cup \{\mu\}}e^{\frac{A_{\sigma}-A_{\mu}}{\eta}}},
\label{eq:probabilities}
\end{equation}
where the denominator is a normalization factor and $\Lambda_\mu$ refers to the
set of sites
adjacent to $\mu$. A vanishing noise intensity $\eta$ yields
the same
issues as the 
deterministic algorithm, while high $\eta$ values produce very loose
pedestrian
configurations; the 
selected intensity, $\eta=1$, offers a good compromise between these extreme
cases.

%\subsection{Resolution of conflicts}

Once the desired sites are selected, distinct pedestrians may aim for the same site. Because this 
site cannot accommodate more than one agent, this generates a conflict, and the physical contacts 
that it involves (``friction'' in the terminology of \cite{Kirchner2003friction}) hamper motion. 
For simplicity, we consider a limit of strong ``friction'', in which competitive conflicts are 
always sterile: nobody then moves forwards.

%\subsection{Cooperative and competitive behaviors of the pedestrians}

Our simulations have shown that the foregoing rules do not lead to a
power-law-tailed distribution 
$p\left(\Delta t\right)$, condition (i) above. This aspect turned out to be difficult to reproduce 
with a cellular automaton and convinced us to examine its origin in granular flows more carefully, 
for these systems are better understood and display similar microscopic statistics. In granular 
hopper flows, clogging occurs because of the formation of pressure-bearing structures such as 
arches \cite{To2001jamming,Garcimartin2010shape}.
The large time lapses $\Delta t$ between successive escapes, forming the tail of $p\left(\Delta 
t\right)$, are dominated by the time that vibrations take to break these arches. Lozano and 
co-workers elucidated that the extent of time an arch resists vibrations on average is controlled 
by its weakest point, i.e., the grain which forms the largest angle with its two
neighbors 
in the arch \cite{Lozano2012flow}. Geometry, and more specifically the variable weaknesses of the 
arches depending on their geometry, thus plays a central role. This calls into question the 
relevance of a lattice-based model such as ours. However, we succeeded in taking account of 
these aspects, without introducing computationally costly geometric considerations.

We started by taking up the original distinction proposed in \cite{Bouzat2014game} between 
cooperative (patient) and competitive (impatient or selfish) pedestrians: here,
at each time 
step, either of the behaviors is chosen probabilistically, on the basis of his/her ``propensity to
cooperate'' $\Pi_{i}(t)\in]0,1]$; he/she will behave cooperatively with probability $\Pi_{i}(t)$, 
and competitively with probability $1-\Pi_{i}(t)$, where $\Pi_{i}(t)$ is fixed to its intrinsic 
value $\Pi_{i}^{\mathrm{(intr)}}$ in the absence of inter-pedestrian contagion.
In effect, the two possible behaviors or strategies differ in the tolerance for not moving. Owing 
to their 
drive, competitive pedestrians find the option to stay at their current position $(x_{i},y_{i})$ 
less attractive than cooperators, the difference being set by $\Pi_{i}(t)$:
\begin{eqnarray}
A(x_{i},y_{i}) & \overset{{\scriptstyle \text{impatient}}}{\longrightarrow} & 
A(x_{i},y_{i})+k\ln\Pi_{i}(t)\label{eq:impatient_attractiveness}\\
A(x_{i},y_{i}) & \overset{{\scriptstyle \text{patient}}}{\longrightarrow} &
A(x_{i},y_{i}). \label{eq:patient_attractiveness}
\end{eqnarray}
The constant $k$ is set to 0.5 to get results quantitatively comparable to the experimental 
measurements of \cite{Pastor2015experimental}.
Thus, competitive agents will be more prone to push for one of the neighboring sites than patient 
ones.

At the constriction close to the exit, an agent will therefore only be able to move forwards if, at 
this time step, the neighbors ``accept'' not to attempt a move to the desired site, which depends 
on their propensities $\Pi_{i}(t)$ via
Eqs.~\ref{eq:impatient_attractiveness}-\ref{eq:patient_attractiveness}.
In a granular system, this would tentatively correspond to a situation in
which neighboring grains in the arch would move slightly backwards 
due to the vibration, thus leaving free space to their neighbor.
Last but not least, to account for the heterogeneous resistances of the
``arches'', the intrinsic 
propensities $\Pi_{i}^{\mathrm{(intr)}}$ are randomly (Gaussian) distributed (remember that, 
without contagion, $\Pi_{i}(t)=\Pi_{i}^{\mathrm{(intr)}}$). This disorder is
critical
with respect to law (i).

\medskip{}

To sum up, at each time step, 
\begin{itemize}
\item[(1)] all pedestrians start by selecting a target site,

\item[(2a)] if the target site is occupied, the pedestrian just waits,

\item[(2b)] otherwise, he/she moves to it, unless other agents are competing
for it (in which case no one moves). 

\item[(3)] Following this first round of motion, some sites have been freshly
vacated, which may allow waiting pedestrians to move to their target
site. Steps (2) are thus iterated until all possibilities of motion
have been exhausted. 
\end{itemize}

The iterative rule (3) allows the formation of files of moving pedestrians,
without voids, and avoids 
an artificial pulsating dynamics. Note that agents cannot move more than once during
a time step (tentatively corresponding to a fraction of a second in reality).

This completes the description of the contagion-free model, that is, the
model featuring fixed propensities
to cooperate $\Pi_{i}(t)=\Pi_{i}^{\mathrm{(intr})}$ for all agents \emph{i}).
Before introducing 
contagion, we dedicate the next section to the study of 
the fixed-behavior model, which reflects the situation expected in controlled evacuations with
prescribed behaviors \cite{Pastor2015experimental}.

\section{IV. Main results for the contagion-free model}

Though the foregoing rules seem rather sensible, it is not granted that they 
can reproduce the microscopic statistical laws (i) and (ii).  Here, we compare
the numerical simulations of the contagion-free model with
experimental data on pedestrian and sheep 
passages through a narrow door 
\cite{Zuriguel2014clogging,Garcimartin2014experimental,Pastor2015experimental,
garcimartin2015flow}.

In Ref.~\cite{Pastor2015experimental}, three levels of
pedestrian cooperativeness were tested: very competitive , moderately
competitive, and cooperative. We arbitrarily define the corresponding
distributions of propensities 
$\Pi^{\mathrm{(intr})}$ as Gaussian distributions of standard deviation 0.2, peaked at $\Pi_{0}=$
 0, 0.4, and 0.8, respectively, and truncated to the interval
{]}0,1{[}, but the results 
are qualitatively robust to variations of these specific values.

%\subsection{Macroscopic observations}
\smallskip

The simulations show close agreement with experimental data, reflected in the
following 
aspects. During the simulated evacuation, the  crowd adopts a semi-elliptic
configuration in 
front of the exit, in broad agreement with experimental observations.
As expected, the mean global evacuation time increases monotonically with the number $N$ of agents. 

As in \cite{Pastor2015experimental}, a ``faster-is-slower'' effect 
\cite{helbing2000simulating,parisi2007faster} is observed, insofar as the evacuation takes longer 
for increasing competitiveness of the agents, i.e., going from cooperative
to moderately and then strongly competitive agents. From now on, the
initial
pedestrian density will
always be set to $\rho=0.6$.

%\subsection{Distribution of time intervals between successive escapes}

\begin{figure}
\begin{centering}
\includegraphics[width=\columnwidth]{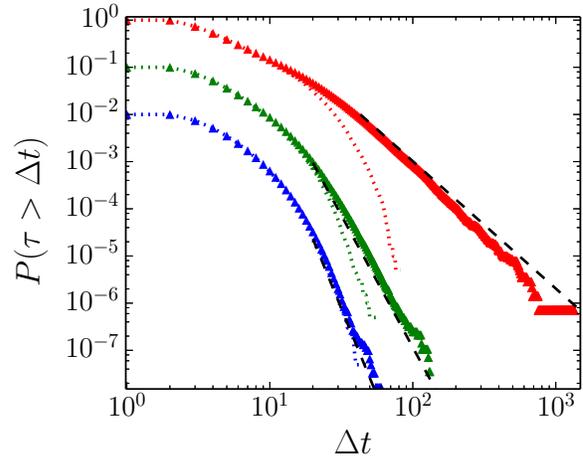}
\end{centering}

\caption{ (Color online) Survival functions $P(\tau>\Delta t)$, for $L_{d}=1$
and different
cooperativeness levels: (red) strongly competitive, (green) moderately
competitive, (blue) cooperative, from top to bottom. The curves have been
shifted vertically to
improve the visibility. The dashed black lines are power-law fits, with the
exponents indicated in Table~\ref{tab:exponent_values}. The dotted lines
are the survival functions obtained by replacing the distribution of
propensities with Dirac
functions peaked at their mean values.}
\label{fig:dist_Deltat_no_contagion_Pi0}
\end{figure}

\begin{figure}
\begin{centering}
\includegraphics[width=\columnwidth]{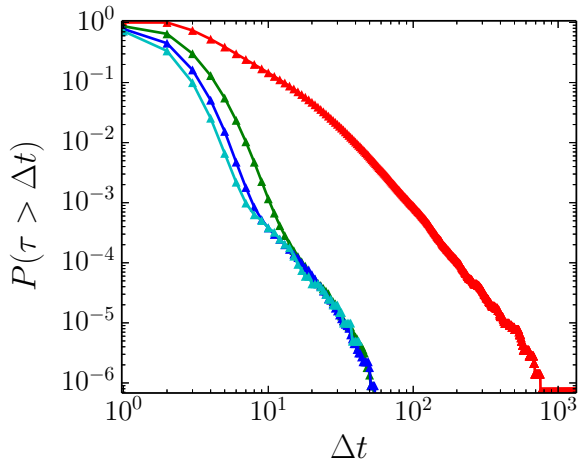}
\end{centering}

\caption{Survival functions $P(\tau>\Delta t)$, for different door widths
$L_{d}=$1, 2, 3, 4, from top to bottom,
with a strongly competitive crowd.}
\label{fig:dist_Deltat_no_contagion_Ld}
\end{figure}

\smallskip
For a more thorough analysis, we compute the microscopic statistics of the evacuation, with the 
help of the Python routine implemented by Alstott et al. \cite{alstott2014powerlaw} on the basis of 
the power-law analysis methods collated by Clauset et al. \cite{clauset2009power}.
Starting with point (i) above,
Figs.~\ref{fig:dist_Deltat_no_contagion_Pi0}-
\ref{fig:dist_Deltat_no_contagion_Ld} present the complementary 
cumulated distribution (or ``survival function'') of time lapses $\Delta t$, in
various 
conditions. Overall, the graphs look similar to the experimental ones 
\cite{Pastor2015experimental}. For the narrowest door, their tails are well
fitted by power laws 
\begin{equation}
p(\Delta t)\sim\Delta t^{-\alpha},\,\alpha>3.\label{eq:power_law_dist}
\end{equation} Indeed, for $L_{d}=1$, the goodness of the power-law fit 
vs. the exponential one is always positive, meaning that the power law provides a better 
fit, with significance values $\mathrm{p}\leqslant 0.05$ (for about $10^6$
sample points); for strongly competitive crowds, $\mathrm{p}$ reaches truly
vanishing values, which makes extremely unlikely the possibility that the
success of the power-law fit over the exponential one is due to chance. These
heavy tails are largely due to
the disorder in the propensities $\Pi_{i}^{\mathrm{(intr)}}$: if the
distributions of 
$\Pi_{i}^{\mathrm{(intr)}}$ are replaced by a Dirac peak at their mean value, the power-law fit 
becomes much poorer (see Fig.~\ref{fig:dist_Deltat_no_contagion_Pi0}).
Appendix~B presents an analytical endeavor towards an approximate
derivation of the microscopic distribution $p(\Delta t)$ for the strongly
competitive crowd, which predicts a power law with exponent $\alpha=4$ (to be
compared to the numerical value $\alpha=3.7$).

%\subsubsection{Influence of cooperativeness}

Large time lapses $\Delta t$ become more frequent for more competitive crowds, which is reflected 
by a lower exponent $\alpha$, consistently with the experimental observations. The model parameters 
were chosen in such a way that the values of the exponents $\alpha$ tend to
match those of 
Ref.~\cite{Pastor2015experimental} for a $70\mathrm{cm}$-wide door, which we here consider to be 
between $L_{d}=1$ and $L_{d}=2$, and perhaps closer to $L_{d}=1$. The (approximate) fitted values
of the exponent are presented in Table~\ref{tab:exponent_values}.

\begin{table}
\begin{centering}
\begin{tabular}{|c|c|c|c|}
\hline 
 & $L_{d}=1$ & $L_{d}=2$ &
Ref.~\cite{Pastor2015experimental}\tabularnewline
\hline 
\hline 
strongly competitive & 3.7 & 4.3 & 4.2\tabularnewline
\hline 
moderately competitive & 6.6 & 3.8(?) & 5.5\tabularnewline
\hline 
cooperative & 8.4 & 5.6(?) & 6.8\tabularnewline
\hline 
\end{tabular}
\par\end{centering}

\caption{Values of the fitted exponents $\alpha$. The values followed by (?)
are strongly dependent on the portion of the curve that is fitted and are
therefore uncertain.}
\label{tab:exponent_values}
\end{table}

%\subsubsection{Influence of the door width}

Figure~\ref{fig:dist_Deltat_no_contagion_Ld} shows the influence of
the door width. Widening the door reduces the probability of long clogs, in
agreement with simulations and experiments on sheep \cite{garcimartin2015flow},
as well as with the intuition. In addition, for door widths $L_d\geqslant2$, the
quality of the fit of the tail with a single power law decreases. Moreover,
while a dramatic change is observed as $L_d$ increases from 1 to 2,
variations in the distribution $p(\Delta t)$, for $\Delta t\geqslant 1$, are
much more tenuous for $L_d\geqslant2$. This does not imply that the
outflow rate is then independent of the door width: simultaneous escapes $\Delta
t=0$ will play a more and more important role as $L_d$ increases. Incidentally,
note
that an apparent insensitivity to large door widths was also reported in 
\cite{song2016selfishness}.

%\subsubsection{Finite-size effects}
The implemented dynamical rules are strictly local, so we expect the escape dynamics at the door to 
be mostly insensitive to finite-size effects. Indeed, for system sizes
$L\geqslant20$ (with initial 
density $\rho=0.6$), the collected ``microscopic'' statistics are virtually independent of $L$. For 
$L<20$, variations are perceptible, because the initial phase in which agents run towards the 
(still uncongested) door is no longer negligible.

%\subsection{Size of the bursts}

\begin{figure}
\begin{centering}
\includegraphics[width=\columnwidth]{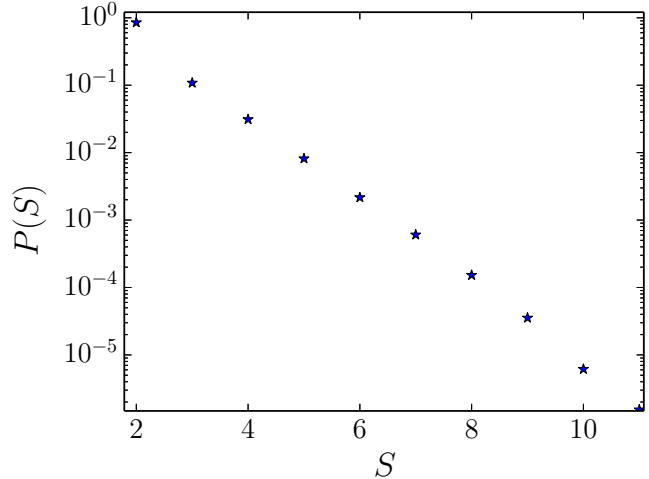}
\end{centering}
\caption{Distribution of burst sizes for
a door width $L_{d}=2$ and a strongly competitive crowd.}
\label{fig:dist_bursts_no_contagion}
\end{figure}

The large fluctuations in the time lapses $\Delta t$ are suggestive of
intermittent dynamics, in 
which bursts of escapes, defined as successive egresses separated by 
$\Delta t\leqslant1$, alternate with long waiting times. In some
experimental settings 
\cite{Zuriguel2014clogging}, these bursts of escapes were found to be
distributed exponentially, see law (ii). In our cellular 
automaton, bursts of escapes (comprising more than one egress) are observed only for $L_{d}>1$ (the 
escape at $L_{d}=1$ is so sluggish that successive escapes are highly improbable); for $L_{d}=2$, 
for instance, Fig.~ \ref{fig:dist_bursts_no_contagion} demonstrates that the
bursts of escapes do indeed follow an exponential law. Note that this law is
more readily obeyed in cellular automata
than law (i), though not systematically \cite{perez2002streaming}.

\paragraph{Micro-macro relation.}

\begin{figure}
\begin{centering}
\includegraphics[width=\columnwidth]{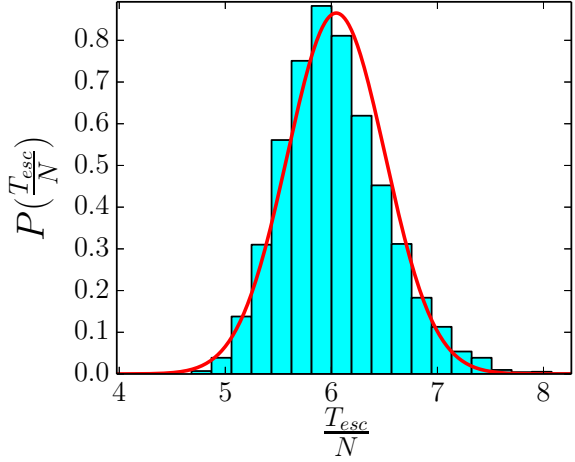}
\end{centering}
\caption{Distribution of global evacuation times in the absence of contagion,
for a strongly competitive crowd, $L_d=1$, and $L=25$. The solid red curve is
the
Gaussian distribution predicted on the basis of Eq.~\ref{eq:P_N_Gaussian}.}
\label{fig:dist_evac_no_contagion}
\end{figure}

Since the time lapses $\Delta t$ fluctuate, the global evacuation time, which
is a sum of time lapses, will also vary between realizations. These variations
between realizations materialize in the distribution 
$P_{N}\left(T_{\mathrm{esc}}\right)$ of global evacuation times of $N\gg1$
agents, an example of which is shown in Fig.~\ref{fig:dist_evac_no_contagion}
for a competitive crowd evacuating through a narrow door. To test
whether the statistical fluctuations can be deduced
from
$p\left(\Delta 
t\right)$ using the micro-macro relation, we compare the actual
histogram of durations $T_{\mathrm{esc}}$ with the Gaussian distribution
predicted on the basis of Eq.~\ref{eq:P_N_Gaussian} and $p(\Delta t)$. 
The excellent agreement that is obtained validates the micro-macro
relation, as expected in a situation without long correlations in the
successive $\Delta t$. On a side note, let us illustrate the
importance of heeding the fluctuations in the distribution, and not
just the average. Suppose that safety standards were set by considering only the
mean
evacuation time; as a
precaution, this mean value could be inflated by, say,
10\% to set the norm. Still, in the particular example of
Fig.~\ref{fig:dist_evac_no_contagion}, actual evacuations would take longer than
the norm in about 8\% of the realizations, even though the global evacuation
time is a statistical average over $\rho L^2\approx 400$ evacuees.

\section{V. Impact of contagion}

In the previous sections, we introduced a minimal model based on a
computationally highly 
efficient cellular automaton capable of reproducing important experimental evacuation data 
semi-quantitatively and we showed that its results obeyed the micro-macro
relation. However, so far, 
the pedestrian behaviors have been kept fixed, with no possibility of change due
to social contagion. This deficiency is remedied in this section, and we
demonstrate that the presence of contagion can lead to the violation of the
micro-macro relation.

In realistic conditions, the propensities to cooperate
$\Pi_{i}(t)$ will 
vary with time, in particular under the negative influence of the aggressive
acts witnessed in 
one's vicinity: having a neighbor of ours choose a competitive
strategy reduces our 
propensity to cooperation in the future; the reverse effect, that is, the positive influence of 
cooperative acts, is deemed weaker and is disregarded here. 
It is worth remarking that the existence of such contagion is endorsed by many social psychology 
theories, even when they claim that cooperativeness often prevails. For instance, the experimental 
results of Kugihara \cite{kugihara2001effects} supported the social identity
model according to which (even in evacuation conditions) people do not break
free of social constraints but conform to the salient local norms; but,
more precisely, they ``showed that what directly affects norm formation
is 
the density of stimulus, that is, the amount of aggression received from others and of 
others\textquoteright{} escape activity divided by group size.''

%\subsection{Contagion dynamics}

\smallskip
Contagion occurs only among neighbors. In the model, it shall be implemented by
means 
of an equation of the form
\begin{equation}
\tau_{\mathrm{rel}}\frac{\mathcal{D}\Pi_{i}}{\mathcal{D}t}(t)=-\underset{\text{relaxation}}{
\underbrace{\left[f\left(\Pi_{i}\right)-f\left(\Pi_{i}^{\mathrm{(intr)}}\right)\right]}}-\underset{
\text{contagion}}{\underbrace{JD_{i}(t)}},\label{eq:contagion_cont}
\end{equation}
where $\nicefrac{\mathcal{D}}{\mathcal{D}t}$ denotes a discrete time derivative (to be specified 
later), $\tau_{\mathrm{rel}}$ sets the duration of the memory, $J$ is the contagion strength, 
$D_{i}(t)\leqslant4$ is the number of neighbors of \emph{i} who chose a
competitive strategy at 
time step $t$, and the function $f$ controls the return 
force of the propensity to collaborate $\Pi_i(t)$ to its intrinsic value
$\Pi_i^{\mathrm{(intr)}}$. To make the boundaries at $\Pi=0$ or 1 
strongly repulsive and introduce nonlinearities in the equation, we set
\begin{equation}
f\left(\Pi\right)\equiv\tan\left[\pi\left(\Pi-\nicefrac{1}{2}\right)\right].\label{eq:f_def}
\end{equation}
We should mention a technical
aspect associated with the discrete time derivative in Eq.~\ref{eq:contagion_cont}:
defining it as a finite difference would require an excessively fine time
discretization to keep $\Pi_{i}$ between 0 and 1. Consequently, we
changed variables to $\psi_{i}=f\left(\Pi_{i}\right)$, wrote 
$\frac{\mathcal{D}\Pi_{i}}{\mathcal{D}t}(t)=\frac{1}{f^{\prime}\left(\Pi_{i}\right)}\frac{d\psi_{i}}
{dt}$,
and set an arbitrary upper bound on the derivative $f^{\prime}$ to impede
too sudden variations of $\psi_i$; $\Pi_i(t+1)$ is then calculated as
$f^{-1}(\psi_i(t+1))$. This scheme allowed us to use the same discretization
time step $\delta t=1$
for the behaviors and for the motion.

Clearly, the foregoing choices are arbitrary to a large extent and we do not claim that 
Eq.~\ref{eq:contagion_cont} precisely describes a real contagion process; yet,
we shall see that it 
yields an interesting phenomenology with plausible practical relevance.

To better understand the implications of Eq.~\ref{eq:contagion_cont}, consider the case in which 
one agent \emph{j} is suddenly struck with panic, so that $\Pi_{j}(t)$ reaches a value $c\ll1$. 
Then, if the crowd is highly susceptible, with $J\rightarrow\infty$, the
nervous or impatient
 behavior of this agent will contaminate others, so that
(in this limit of strong contagion) the whole crowd may turn nervous.

\subsection{Analytical approach}
\smallskip
We now turn to a more quantitative analysis of the contagion dynamics.
The analytical study is premised on a quasi-equilibrium assumption:
the evacuation is considered slow enough for the psychological propensities
to reach their equilibrium values before any significant change in
the geometric configuration of the crowd. A similar assumption was
used in Ref.~\cite{Heliovaara2013patient}. Under this hypothesis,
in the stationary state, we can average Eq.~\ref{eq:contagion_cont}
over a reasonably large number of time steps, viz.,
\begin{eqnarray}
\tau_{\mathrm{rel}}\overline{\frac{\mathcal{D}\Pi_{i}}{\mathcal{D}t}}(t) & = & 
-\left[\overline{f\left(\Pi_{i}\left(t\right)\right)}-f\left(\Pi_{i}^{\mathrm{(intr)}}\right)\right]
-J\overline{D_{i}(t)},\label{eq:eq_contag_stat1}  \nonumber \\
&&
\end{eqnarray}
where the overbars $\overline{\bullet}$ denote time-averaged
quantities.

Remarking that $D_{i}(t)=\sum_{j \in \Lambda_i}d_{j}(t)$,
where $d_{j}(t)=1$ if agent $j$ behaved competitively at time step
\emph{t} and 0 otherwise and $\Lambda_i$ denotes the first neighbors
of $i$, we see that $\overline{D_{i}(t)}=\sum_{j \in
\Lambda_i}\left[1-\overline{\Pi_{j}(t)}\right].$

\paragraph{Rewriting with a potential.}
\vspace{2mm}

To leading order in $\left(\Pi_{i}\left(t\right)-\overline{\Pi_{i}(t)}\right)$,
$\overline{f\left(\Pi_{i}\left(t\right)\right)}\approx 
f\left(\overline{\Pi_{i}\left(t\right)}\right)$
and Eq. ~\ref{eq:eq_contag_stat1} can then be recast into 
\begin{eqnarray*}
\tau_{\mathrm{rel}}\overline{\frac{\mathcal{D}\Pi_{i}}{\mathcal{D}t}}(t) & = & 
-\frac{dV_{i}}{d\overline{\Pi_{i}}}\left(\overline{\Pi_{i}}\right)-\sum_{j \in
\Lambda_i 
}\frac{dV^{(2)}}{d\overline{\Pi_{i}}}\left(\overline{\Pi_{i}},\overline{\Pi_{j}}\right).
\end{eqnarray*}
where we have introduced the potentials
\[
V_{i}(\Pi)\equiv F(\Pi)-\Pi 
f\left(\Pi_{i}^{\mathrm{(intr)}}\right)+z_{i}J\left(\Pi-\frac{1}{2}\Pi^{2}\right)
\]
\[
V^{(2)}\left(\Pi_{i},\Pi_{j}\right)\equiv\frac{J}{2}\left(\Pi_{i}-\Pi_{j}\right)^{2},
\]
$F$ is a primitive of $f$, e.g., 
$F(\Pi)\equiv-\frac{1}{\pi}\ln|\cos\left[\pi\left(\Pi-\nicefrac{1}{2}\right)\right]|$
and $z_{i}$ is the number of first neighbors of agent \emph{i}.
The pair potential $V^{(2)}$ is symmetric and is interpreted as an
interfacial cost.

The total energy of the system is then
\[
\mathcal{V}\left(\{\overline{\Pi_{i}}\}\right)=\sum_{i}V_{i}(\overline{\Pi_{i}}
)+ \frac{1}{2}\sum_{ (i, j) | j
\in
\Lambda_i }V^{(2)}\left(\overline{\Pi_{i}},\overline{\Pi_{j}}\right).
\]

%\subsubsection*{Mapping onto an Ising system in given limits}

\begin{figure}
\begin{centering}
\includegraphics[width=\columnwidth]{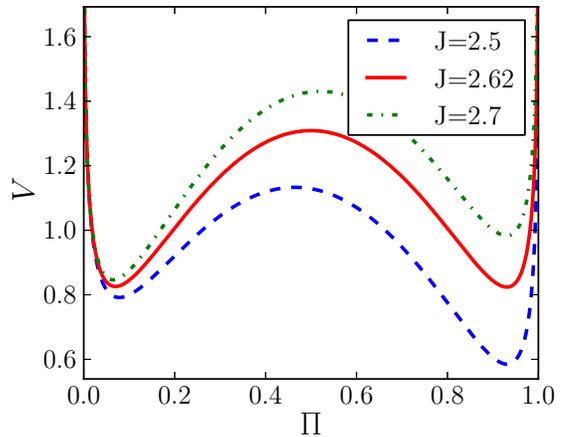}
\end{centering}
\caption{Potential $V$ for $\Pi^{\mathrm{(intr)}}=0.93$,
$z=4$, and different coupling parameter $J$, as indicated in the
legend.}
\label{fig:Potential_V}
\end{figure}

If one neglects heterogeneities in the intrinsic 
distributions ($\Pi_{i}^{\mathrm{(intr)}}\rightarrow\Pi^{\mathrm{(intr)}}$) and in the 
configuration ($z_{i}\rightarrow z$), $V_{i}$ becomes independent of \emph{i} ($V_{i}\rightarrow 
V$), and the ground state $\left\{ \Pi_{j}\right\} $ for the energy is obtained for the homogeneous
system at $\Pi_{j}=\Pi$, where $\Pi$ minimizes $V$, viz., $\nicefrac{dV}{d\Pi}=0$. 
Interestingly, for some values of the coupling parameter $J$ and $\Pi^{\mathrm{(intr)}}$, the 
potential $V$ displays bistability (see
Fig.~\ref{fig:Potential_V}), with two distinct minima at
$\Pi_{\downarrow}$ and 
$\Pi_{\uparrow}$ such that, by 
definition, $V\left(\Pi_{\downarrow}\right)<V\left(\Pi_{\uparrow}\right)$.  For
$z=4$ and $\Pi^{\mathrm{(intr)}}=0.94$, the minima are approximately of equal
depths when $J\simeq2.62$.

Since, at not too high temperature, all $\Pi_{j}$s will dwell close
to the bottom of an energy basin, it is convenient to write $\Pi_{j}=\Pi_{\downarrow}+\tilde{\Pi}$
as $\left(\downarrow,\tilde{\Pi}\right)$ (or $\Pi_{j}=\Pi_{\uparrow}+\tilde{\Pi}$
as $\left(\downarrow,\tilde{\Pi}\right)$) in the bistable state. If the thermal
deviations $\tilde{\Pi}$ are overlooked, the model
can directly be mapped onto a two-dimensional (2D) Ising model in an external
field
$h$, with the following Hamiltonian

\[
\mathcal{H}=-h\sum_{i}\sigma_{i}- \frac{J_{\mathrm{Ising}}}{2}\sum_{ (i, j) | j
\in
\Lambda_i }\sigma_{i}\sigma_{j},
\]
and $\sigma_{i},\sigma_{j}\in\left\{ -1,1\right\} $, with 

\[
\begin{cases}
h & =\frac{V\left(\Pi_{\downarrow}\right)-V\left(\Pi_{\uparrow}\right)}{2}\\
J_{\mathrm{Ising}} & =\frac{J}{4}\left(\Pi_{\uparrow}-\Pi_{\downarrow}\right)^{2}
\end{cases}
\]
It is well known that, in the absence of an external
field $h$, the 2D Ising system undergoes a phase transition from a disordered
state with mixed spins to an ordered ($\uparrow$ or $\downarrow$)
state as the temperature declines and that the transition is associated
with diverging correlation lengths. The thermodynamic transition disappears
at $h\neq0$; nevertheless, a vestige of the criticality persists at small but
finite $h$.

%\subsubsection*{Estimation of temperature}

There are no \emph{bona fide }thermal fluctuations in
Eq.~\ref{eq:contagion_cont},
but the instantaneous discrepancies between $d_{j}(t)\in\left\{ 0,1\right\} $
and $1-\overline{\Pi_{j}(t)}$ introduce fluctuations $\xi(t)$ in
practice, viz.,
\[
\tau_{\mathrm{rel}}\frac{\mathcal{D}\Pi_{i}}{\mathcal{D}t}(t)=-\frac{dV_{i}}{d\overline{\Pi_{i}}}
\left(\overline{\Pi_{i}}\right) - \frac{1}{2} \sum_{(i, j) | j
\in
\Lambda_i
}\frac{dV^{(2)}}{d\overline{\Pi_{i}}}\left(\overline{\Pi_{i}},\overline{\Pi_{j}}\right)+\xi(t).
\]
By comparing the first and second moments of $\xi$, namely, $\left\langle \xi\right\rangle =0$
and $\left\langle \xi{}^{2}\right\rangle 
\approx4J^{2}\overline{\Pi_{i}}\left(1-\overline{\Pi_{i}}\right)$,
with the thermal relation $\left\langle \xi{}^{2}\right\rangle =2\tau_{\mathrm{rel}}T$,
we get a lower bound for the effective temperature, 
$T_{\mathrm{eff}}=\frac{2J^{2}}{\tau_{\mathrm{rel}}}\overline{\Pi_{i}}\left(1-\overline{\Pi_{i}}
\right)$.

\subsection{Numerical results}

%\subsubsection{Spatial organization of the crowd}

\begin{figure*}
\begin{centering}
\subfloat[$J=2.96$]{\includegraphics[width=6cm]
{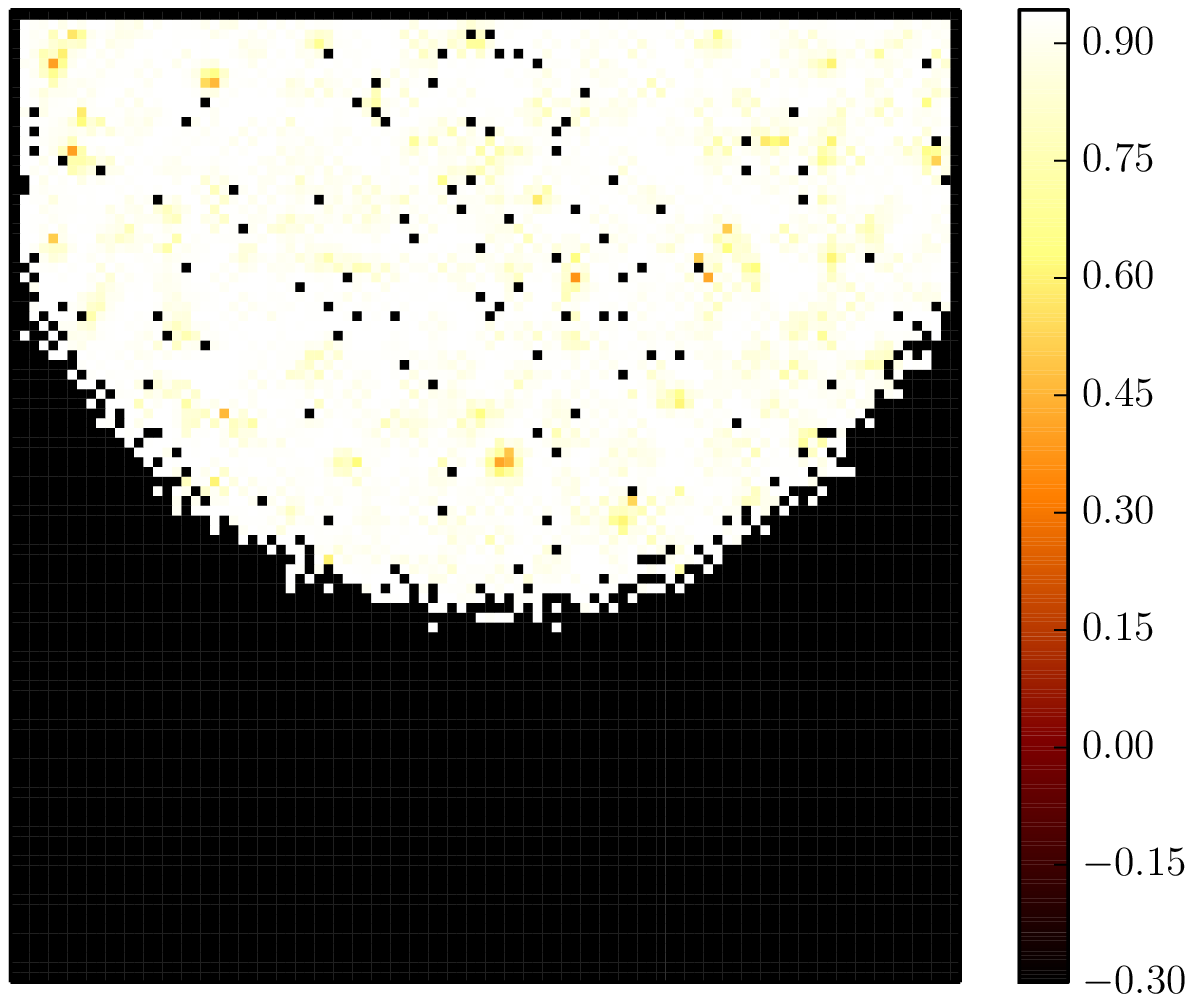}}
\subfloat[$J=2.9704$]{\includegraphics[width=6cm]
{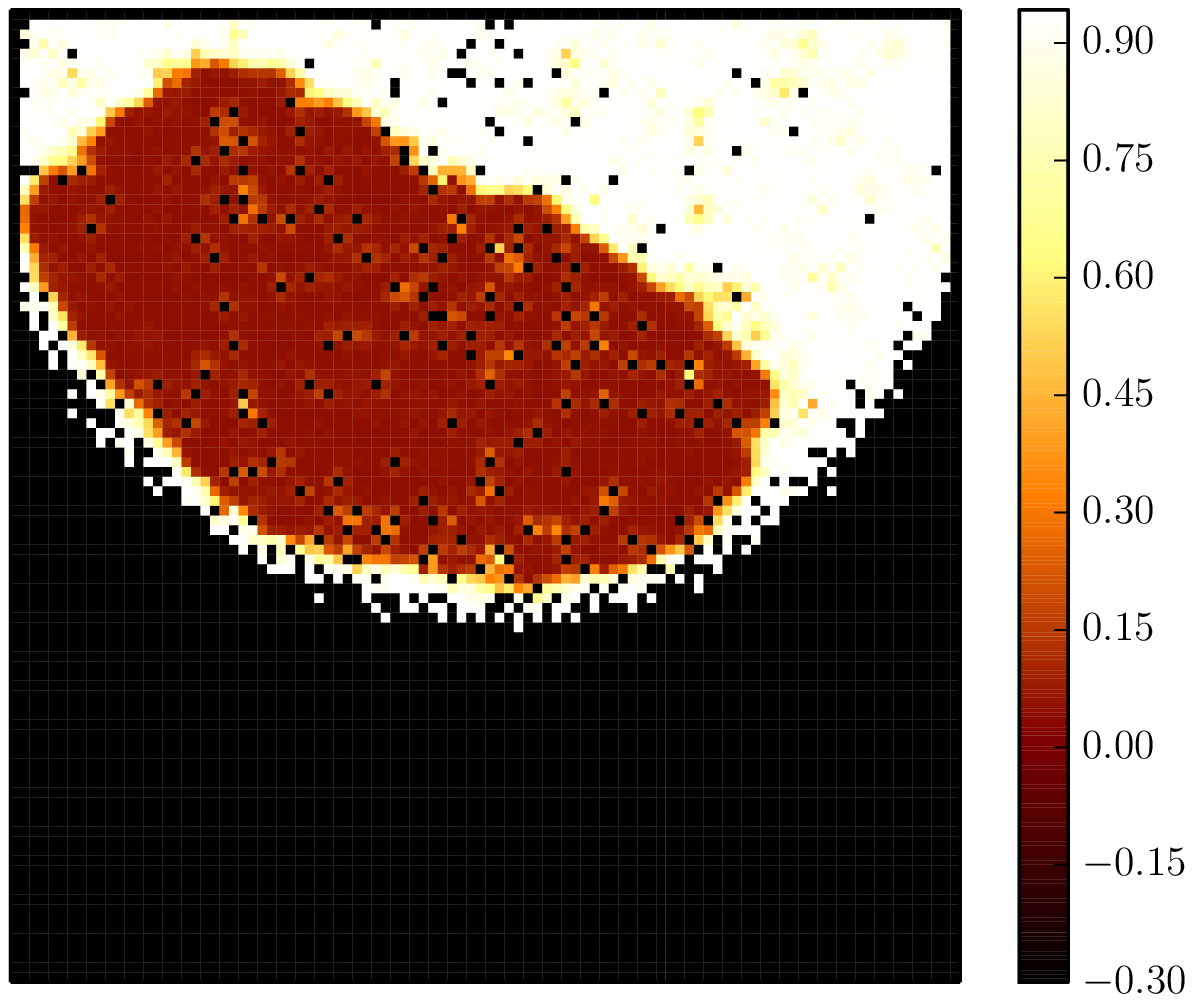}}
\subfloat[$J=2.98$]{\includegraphics[width=6cm]
{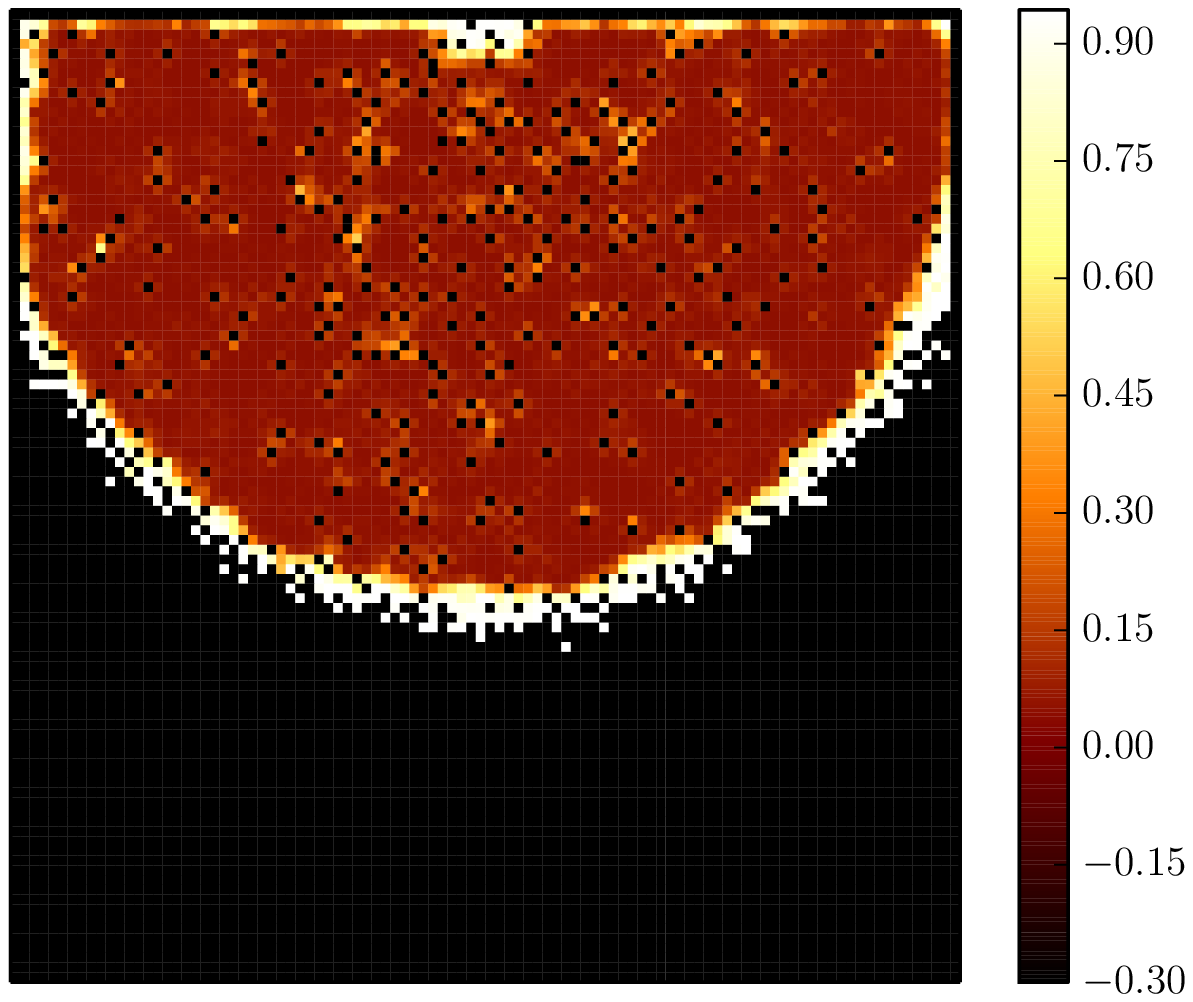}}
\end{centering}

\caption{(Color online) Color maps of the propensities $\Pi_{j}(t)$
in the crowd at time $t=4000$, for $\Pi^{\mathrm{(intr)}}=0.94 \pm 10^{-7}$.
Vacant sites appear in black.}
\label{fig:PropensityMaps}
\end{figure*}

Numerical simulations of the model confirm the validity of the foregoing
analysis. Snapshots of the propensities of the evacuating crowd are shown in
Fig.~\ref{fig:PropensityMaps} as color maps, for different values
of the coupling parameter $J$. Clearly, as $J$ is varied across
a critical value $J^\star$, large domains
of strongly correlated propensities are observed in the system, which
is suggestive of the close presence of a critical point. Two additional
spatial feature are noticeable. There is an abundance of cooperative
(high-$\Pi$)
agents in the
vicinity of the exit, along with a cooperative fringe at the outer edge of
the crowd, where each agent has fewer
neighbors, i.e., lower coordination ($z$) values, and thus
fewer contagion possibilities.

These results hold only for a sharply peaked distribution of intrinsic
propensities $\Pi^{\mathrm{(intr)}}$; for more broadly distributed
$\Pi^{\mathrm{(intr)}}$ (with standard deviations larger than, say,
0.01),
the level of disorder increases and the spatial correlations of the
propensity become less visible at $J^\star$.

%\subsubsection{Finite-size effects}

As the size $L$ of the system  decreases, the border and vacant sites become
proportionally more important, so that the critical contagion strength
$J^{\star}(L)$ at the onset of the evacuation shifts to larger values; for
example, we observed
that $J^{\star}(L=100)\approx2.97$, $J^{\star}(L=40)\approx3.11$,
and $J^{\star}(L=25)\approx3.33$. This notably implies that the critical value
$J^\star$ will increase during the evacuation as fewer and fewer people are left
in the room.

%\subsubsection{Variability between realizations and the violation of the micro-macro relation}
\smallskip

Let us now move on to the possible effect of the contagion rule on
the evacuation.

The existence of large correlated domains in the unbounded system
implies that, should the system be small enough, i.e., of order
the correlation length or below, there will be occurrences where the
crowd will escape cooperatively and others, where ``panic'' will
spread and foster selfish behavior. We claim that, notwithstanding the
arbitrariness of the chosen rules, this behavioral dichotomy is a practically
relevant
general effect due to social contagion. In particular, taking
it into account may help settle the debate between the staunch supporters
of the maintenance of social cooperation in emergency evacuations
\cite{drury2009everyone,drury2012collective} and the proponents of
the emergence of selfish or aggressive behaviors in emergency, as
often implemented in physics-based models %
\footnote{Note that the effect of cohesion enhancement through identification
within the group predicted the social identity theory would be accounted
for by an increase in $\Pi^{\mathrm{(intr})}$, and possibly by a
dependence of these propensities on the neighbors%
}.

\begin{figure}
\begin{centering}
\subfloat[$J=3.35$ and
$\Pi^{\mathrm{(intr})}=0.94$, system size:  $L=25$.]
{
\includegraphics[width=0.4\columnwidth]{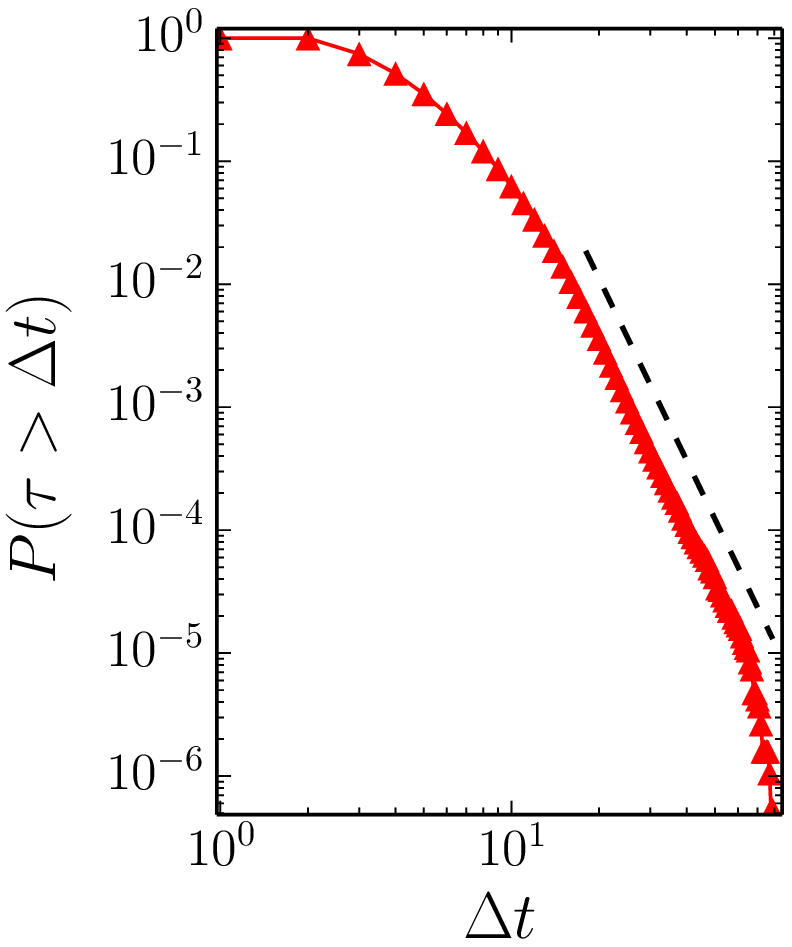}
\includegraphics[width=0.59\columnwidth]{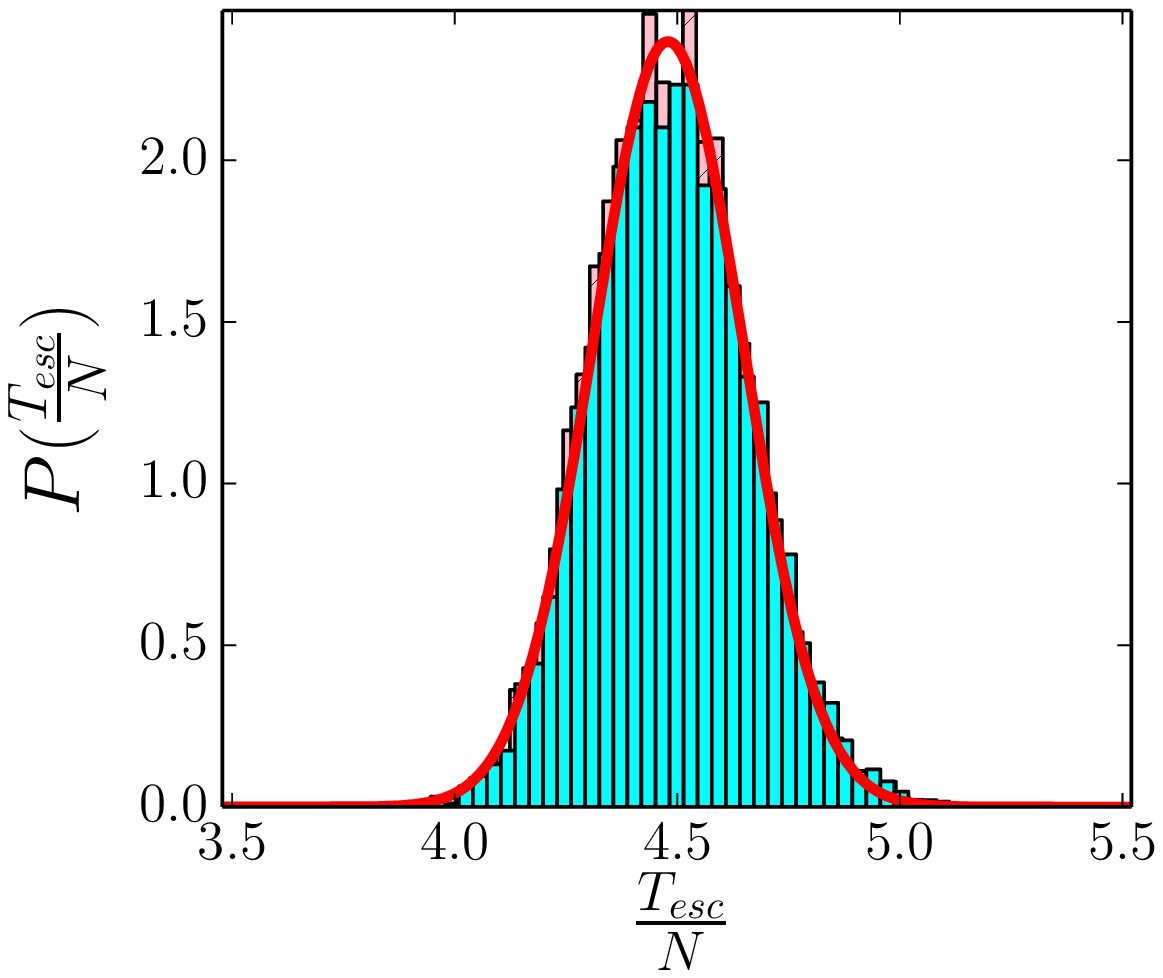}
}

\subfloat[$J=453$ and $\Pi^{\mathrm{(intr})}=0.99905$, system size:  $L=30$.]
{
\includegraphics[width=0.4\columnwidth]{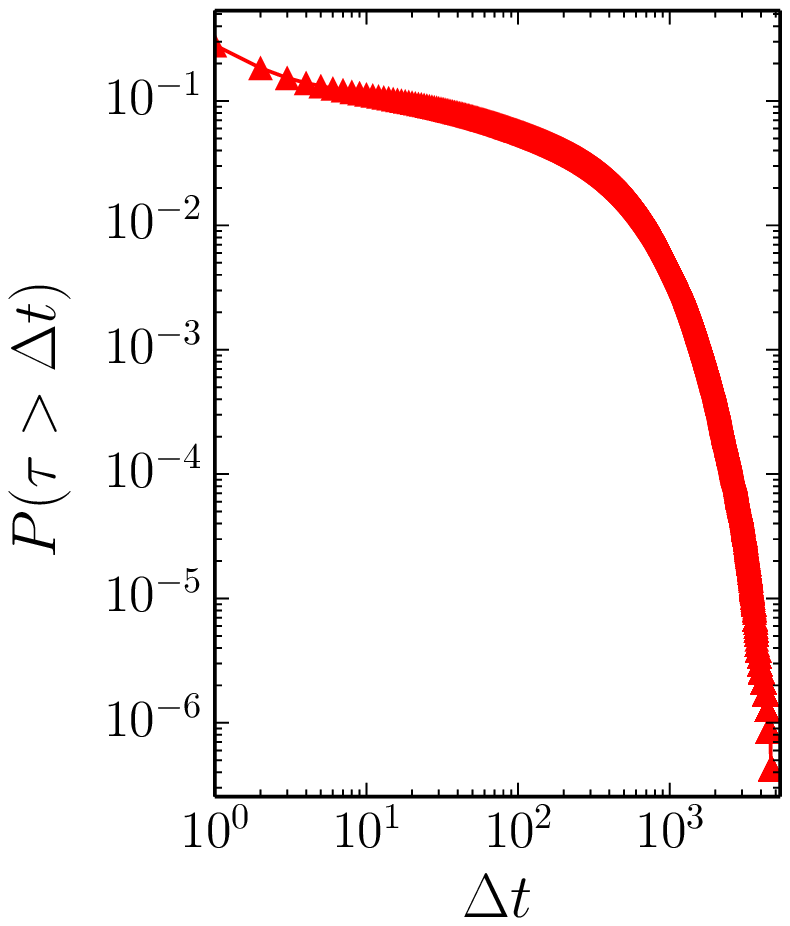}
\includegraphics[width=0.59\columnwidth]{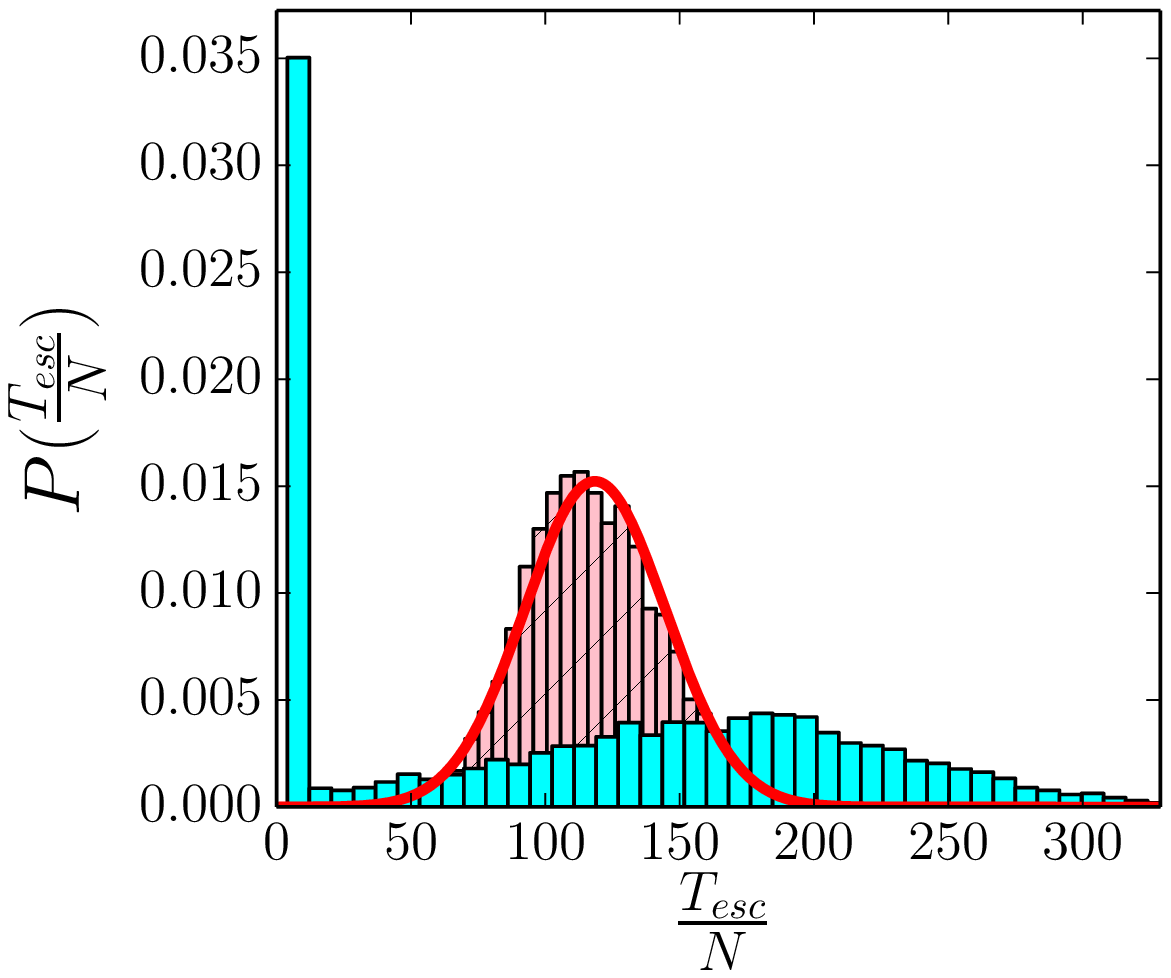}
}

\end{centering}

\caption{(Color online) Evacuations in the presence of behavioral contagion.
Left: survival function. Right: normalized global distribution of evacuation
times per pedestrian $\frac{T_\mathrm{esc}}{N}$, with the simulations in
light blue,
the expectations from the micro-macro relation in hatched pink (see text) and
the
Gaussian distribution of Eq.~\ref{eq:P_N_Gaussian} as a solid red curve.}
\label{fig:micro_macro_cont}
\end{figure}

We turn to the micro-macro relation. The ``microscopic'' distribution
of time lapses $\Delta t$, averaged over many realizations, is plotted
in Fig.~\ref{fig:micro_macro_cont}a for a system close to the critical point,
$J\simeq J^\star(L)$ and $L=25$. Despite the presence of correlated domains
in the simulations (\emph{not shown}), the Gaussian prediction of
Eq.~\ref{eq:P_N_Gaussian} based on the micro-macro relation in the limit
$N\rightarrow \infty$ seems to capture the actual distribution of global
evacuation evacuation times satisfactorily. To test the agreement more
thoroughly, we measured the distance between the distribution of
$N_\mathrm{sim}$ simulated evacuation times and the distribution of
$N_\mathrm{sim}$ sums of $N$ time lapses \emph{randomly} drawn from the
microscopic distribution. As
a statistical measure of distance (or rather proximity) between
finite-size distributions, we chose the p-value associated with either the
Mann-Whitney test or the Kolmogorov-Smirnov test. Despite collecting about
$N_\mathrm{sim}\approx 5000$ global evacuation
times for every set of parameters and averaging over many random distributions,
we did not measure any p-value sufficiently small to ascertain a significant
difference
between the actual distribution and the random one, nor did we observe
variations of the p-value with the contagion strength $J$ that were
substantially above the statistical noise.
Therefore, if the micro-macro relation is violated, the violation is
at most tenuous. This could be due to the finite life-time of the
domains of correlated propensities or to the aforementioned persistence of
cooperative behaviors near the exit (compared to the bulk), where time lapses
are computed.

Thus, deviations from the micro-macro relation in the bistable system
considered above are, at best, moderate. Let us now consider another mechanism
that may induce deviations, namely the scenario (mentioned at the end of Section
II) of
an initially cooperative crowd that is highly
susceptible
to ``panic'', i.e., with very large $J$. In this case, the crowd is initially
trapped in a metastable high-$\Pi$ state and the question is whether,
in a given realization, nervous or impatient behaviors will have time to
nucleate and push the crowd into the stable, low-$\Pi$ state, thus delaying
the evacuation (all the more so as these behaviors have nucleated
early), or not.  Figure~\ref{fig:micro_macro_cont}b presents the microscopic
and macroscopic distributions associated with this scenario. Once again, the
macroscopic distribution is compared with a random distribution inferred from
the microscopic statistics. Here, the discrepancy is blatant, which reflects the
dramatic failure of the micro-macro relation. In particular, the inferred
distribution captures neither the frequency of fast (cooperative) evacuations
nor the occurrence of sluggish evacuations where competitive behaviors
pervaded the crowd.

\section{VI. Discussion and Potential relevance for safety science}

It is a crucial requirement in any safety protocol that buildings should provide
adequate means of egress in the event of an evacuation, by preserving  unlocked,
unobstructed, and clearly marked 
emergency exits. There is a plethora of historical examples where an
insufficient outflow capacity was reported as a decisive factor leading to
a crowd disaster. Indeed, in their haste to exit, pedestrians may tend to push
their neighbors, which can result in the buildup of pressure at the exit and
its congestion. Recent experimental works have raised the hope of a 
quantitative characterization of the outflow capacity of a given exit, based on the distribution of 
time lapses between successive egresses and depending on the eagerness to egress. 

In the present work, we have developed a minimalistic cellular automaton model able to reproduce 
semi-quantitatively the major features observed in (controlled) evacuation
experiments, which is probably a first. In 
particular, for competitive egresses we obtained a distribution exhibiting a
large tail that is well described 
by a power law.
For sure, our model is but a minimal abstraction of the real problem; still, it
is
noteworthy that our approach is in line with more detailed previous
studies that
endeavored to go beyond the reliance of the social force paradigm on
binary interactions between pedestrians. In \cite{moussaid2011simple} the
authors considered that, instead of being repelled
by their neighbors, the individuals look for a free path through the crowd. In
this sense, our cellular automaton is based on rather similar considerations.

Among other features, we focused on the effect of social contagion. As expected,
in its absence, the (practically relevant) ``macroscopic''
distribution of global evacuation times is reliably 
inferred from the ``microscopic'' statistics of time lapses. On the contrary,
should there be
a possibility of behavioral contagion, this micro-macro relation may be violated
due to the induced correlations in the behaviors of the pedestrians within each 
evacuation. While its origin can be attributed to complex causes and its study lies in the realms 
of social psychology, we have abstracted and illustrated this point by implementing simple 
contagion dynamics, which turned out to be quantitatively connected to the 2D
Ising model: in some regions of parameter space, the crowd displays
bistability, with a 
cooperative state and an impatient one. However, large violations of the
micro-macro relation were in fact only observed at very large contagion
strengths, when
the crowd initially resided in a metastable cooperative state.

This observation is conceptually interesting in that it shows that, even in
strictly identical conditions, the durations of evacuation
may vary much more than suggested by naive reasoning (the micro-macro
relation), although numerically this happened only in a select region of
parameter space; the higher variability
was then due to contagion-induced correlations.

In reality, it is highly unlikely that two evacuations take
place in strictly identical conditions: crowds will differ in their composition
and their psychological states will be strongly affected by previous
circumstances, not to mention the singularity of each evacuation and the
various possible emergency stimuli. Within the simple framework of
our model, these differences might be accounted for by variations of both the
distribution of intrinsic probabilities $\Pi^{\mathrm{(intr)}}$
and the contagion strength $J$ with the occurrence. In any event,
it is natural to expect that this enhanced variability between
realizations will result in more systematic deviations from the micro-macro
relation than in our study.

Given this failure of the micro-macro relation, should experts in safety science dedicate any 
attention whatsoever to the microscopic distribution of time lapses (which,
unlike the global one, can potentially be 
compiled)? As a matter of fact, we  believe
that this approach still represents a step forward with respect to
the traditional use of a single flow rate value
(e.g., an exit capacity of 82 persons per meter width per minute, according
to the British Guide to Safety at Sports Grounds \cite{greenguide2008}), in that
it accounts for statistical fluctuations (but not for context-related
variations). 
In this spirit, one may envision a conservative approach to the assessment of
the exit capacities. Regardless of the geometry, it would consist in 
extracting the microscopic distribution from the \emph{slowest} evacuations of a
densely-crowded room, 
from a batch of real video-recorded evacuations, and deriving the global
distribution from the micro-macro relation, as a function of the attendance.
Data from evacuation drills might also be considered, but this demands to find a
trade-off in the dilemma between prescribing realistically competitive
behaviors to the participants and ensuring a safe drill.
Alternatively, the avoidance of worst-case scenarios in real evacuations could
be left to the responsibility of other
mitigating strategies (i.e., not related to the geometry); then, under the
assumption that the evacuating crowd does not yield to panic, typical egresses
could be used to compile the microscopic statistics and derive the exit
capacity. Finally, it is worth recalling that the possibilities 
of delay contemplated here arise because of congestion at the exit and are
naturally be less 
pronounced with wider doors or a scarcer crowd.

On a more general note, our work is yet another example of the potential 
of Statistical physics models to provide a phenomenological account of
society-related topics \cite{bouchaud2013crises}. 
Along this line, we remark that, both in the (Statistical) physics of 
complex systems and in the social psychology of crowds
\cite{le1995psychologie,johnson1977crowd},
many a subtlety of the individual entity is left behind when it forms part of an assembly and the 
collective response may be more polarized than that of the isolated entities.

\acknowledgments
This work received financial support from CONICET (PIP 112-201101-0310),
and Universidad Nacional de Cuyo (06/C304).

\appendix
\section{Appendix A: Limit distributions for power-law exponents $\alpha
\leqslant 3$}

Consistently with empirical measurements of the distribution $p(\Delta t)$ of
time lapses $\Delta t$ between successive egresses, in the main text we have
focused on distributions featuring a power-law tail $p(\Delta t)\sim \Delta
t^{-\alpha}$ with an exponent $\alpha>3$. In that case, the sum of time lapses
converges to a Gaussian distribution, as stated by the central limit theorem.

But what happens if $\alpha
\leqslant 3$? For $2<\alpha
\leqslant 3$, the mean value of the distribution is still well defined, but the
standard deviation diverges, and a L\'evy distribution
is expected for the sum of the time lapses. 

Indeed, according to the developments in \cite{nolan2016stable} and
\cite{gnedenko1968limit}, 
and restricting our attention to the case $2<\alpha\leqslant 3$, 
we find the following result. Let $p(\Delta t)$ behave as
$p(\Delta t)=\lambda \Delta t^{-\alpha}$ for large $\Delta t$ and
let $\mu$ be its mean value. Then, the distribution of 
$T_\mathrm{esc}=\sum_{i=1}^N \Delta t_i$ for large $N$ converges to the L\'evy 
density function 
\begin{equation*}
f(T_\mathrm{esc}|\tilde{\alpha},1,\gamma_N,\delta_N,0),
\end{equation*}
with the notations and parameter prescriptions of
Ref.~\cite{nolan2016stable}, where $\tilde{\alpha}\equiv\alpha-1$
($1<\tilde{\alpha}\leqslant 2$), 
$\delta_N \equiv N\mu+\gamma_N \tan{\frac{\pi \tilde{\alpha}}{2}}$, and
\begin{equation*}
\gamma_N \equiv \left(\frac{\pi \lambda
N}{2\tilde{\alpha}\sin(\frac{\pi\tilde{\alpha}}{2})
\Gamma(\tilde{\alpha})}\right)^{\frac{1}{\tilde{\alpha}}}.
\end{equation*}

Recall that this result is conditioned by the validity of the micro-macro
connection, which may fail in the presence of contagion as discussed in the
main text.

\section{Appendix B: Approximate analytical estimation of the microscopic
distribution}

In the main text, we presented simulations of the evacuation of a
strongly competitive crowd through a narrow door of unit width. Numerically,
the distribution $p(\Delta t)$ of time lapses $\Delta t$ was found to be well
fitted by a power law of exponent $\alpha \approx 3.7$ at large $\Delta t$.
Here, we aspire to support these numerical results with approximate analytical
calculations.

First, the rules of the model imply that the largest time lapses
are obtained when three very impatient agents $i$, $j$, and $k$ ($\Pi_i,\Pi_j,
\Pi_k \ll 1$) compete for the site just in front of the exit. In the worst
cases, the sites behind them are also occupied, which deprives them of the
option of stepping backward. Most probably, the conflict will be resolved only
when two of these agents ``accept'' to stay on their current sites, while the
third one attempts to move forward to the desired site.

For $\Pi_i \ll 1$, considering the probabilities of site selection of
Eq.~\ref{eq:probabilities} and the attractiveness defined in
Eqs.~\ref{eq:impatient_attractiveness}-\ref{eq:patient_attractiveness} with
$k=0.5$ and $\eta=1$, the most probable option that leads agent $i$ to stay
on-site
is not to choose a cooperative strategy, but rather to adopt a competitive
behavior (with probability $1-\Pi_i$) and then select the current site (with
probability $p^s_i$ of order $e^\frac{-1}{\eta}\,e^{\frac{k\ln \Pi_i}{\eta}}
\sim
\Pi_i^\frac{k}{\eta} = \sqrt{\Pi_i}$). To leading order in the propensities
and up
to numerical prefactors, the probability $P(\Delta t=n)$ of observing a time
lapse of duration $n\gg1$ is given by the probability of a succession of $(n-1)$
conflicts
finally resolved by two agents choosing to stay on-site, viz.,

\begin{equation*}
  \int d\Pi_i D(\Pi_i) \int d\Pi_j D(\Pi_j) \int d\Pi_k
D(\Pi_k) \left(1- s^{(2)}\right)^{n-1} s^{(2)},
\end{equation*}
where $D(\Pi_\gamma)$, $\gamma \in \{i,j,k\}$, is the
distribution of propensities and $s^{(2)}$ is a shorthand for $(p^s_i p^s_j +
p^s_j p^s_k + p^s_i p^s_k)$. For the strongly competitive crowd, it
is fair to approximate the truncated Gaussian distribution $D(\Pi_\gamma)$ by a
constant for $\Pi_\gamma < \epsilon$, where $\epsilon$ is a tiny constant.
Changing variables $\Pi_\gamma$ to
$p^s_\gamma \equiv \sqrt{\Pi_\gamma}$, we get
\begin{equation*}
 P(\Delta t=n) \sim \int_0^\epsilon  dp^s_i p^s_i \int_0^\epsilon  dp^s_j p^s_j 
\int_0^\epsilon dp^s_k p^s_k
 \left(1- s^{(2)}\right)^{n-1} s^{(2)}.
\end{equation*}
The triple integral can be split into a region $p^s_k < p^s_j < p^s_i$ and five
other symmetric regions. In the first region, we make
the approximation $s^{(2)} \approx p^s_i p^s_j$, so that, up to numerical
factors, $ P(\Delta t=n) \sim $
\begin{equation*}
\int_0^\epsilon  dp^s_i  \int_0^{p^s_i}  dp^s_j  \int_0^{p^s_j}  dp^s_k 
 \left(1- p^s_i p^s_j\right)^{n-1} {(p^s_i)}^2 {(p^s_j)}^2 p^s_k.
\end{equation*}
Finally, the integrand reaches its maximum at $p^s_i \approx p^s_j \approx p^s_k
\sim n^{\nicefrac{-1}{2}}$ and rapidly decays for larger $p^s_i$, so that,
discarding the part $p^s_i > n^{\nicefrac{-1}{2}}$ and approximating the
integrand by its maximum, we arrive at
\begin{equation*}
  P(\Delta t=n) \sim n^{-4},
\end{equation*}
that is to say, the distribution of time lapses decays as a power law with
exponent $\alpha = 4$. Considering the many inaccuracies involved in the
foregoing calculation, we deem the agreement between the analytical result and
the numerical one ($\alpha \simeq 3.7)$ quite satisfactory.

\section{Appendix C: Power-Law Fits}

\begin{figure}
\begin{centering}
\includegraphics[width=0.48\columnwidth]{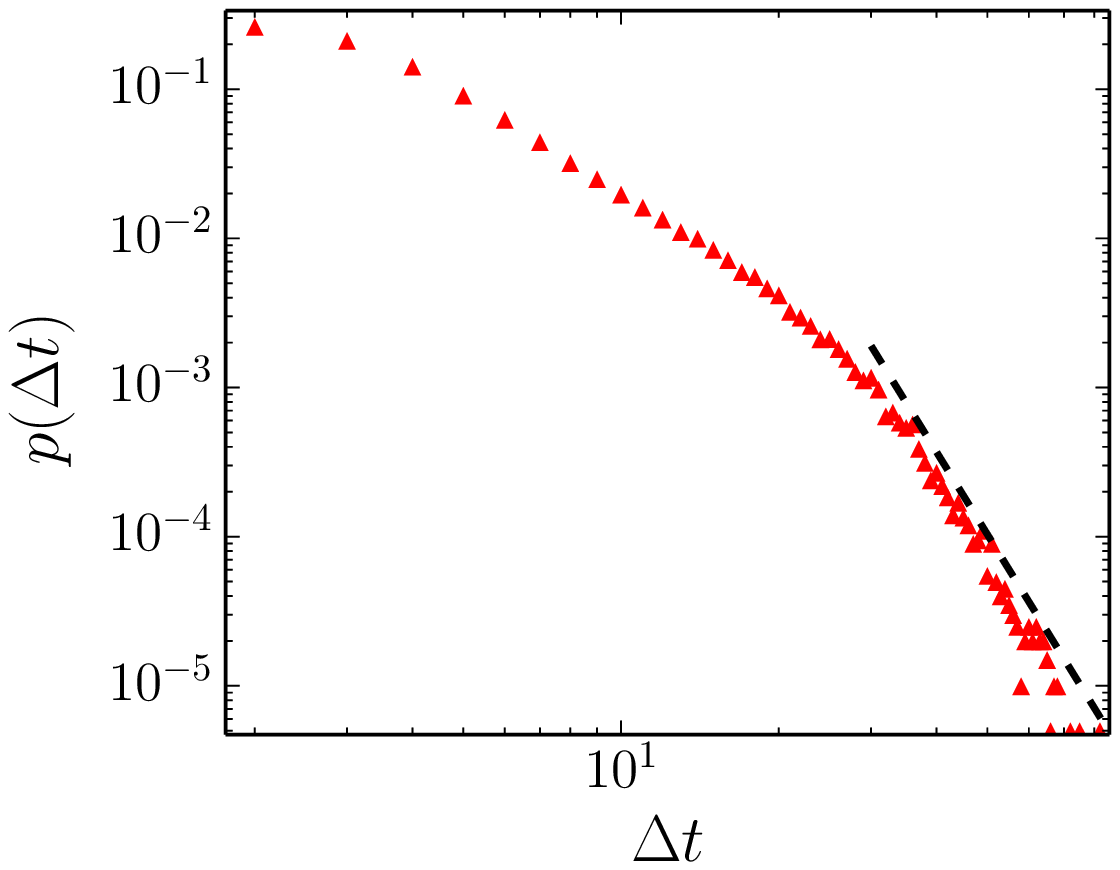} 
\includegraphics[width=0.48\columnwidth]{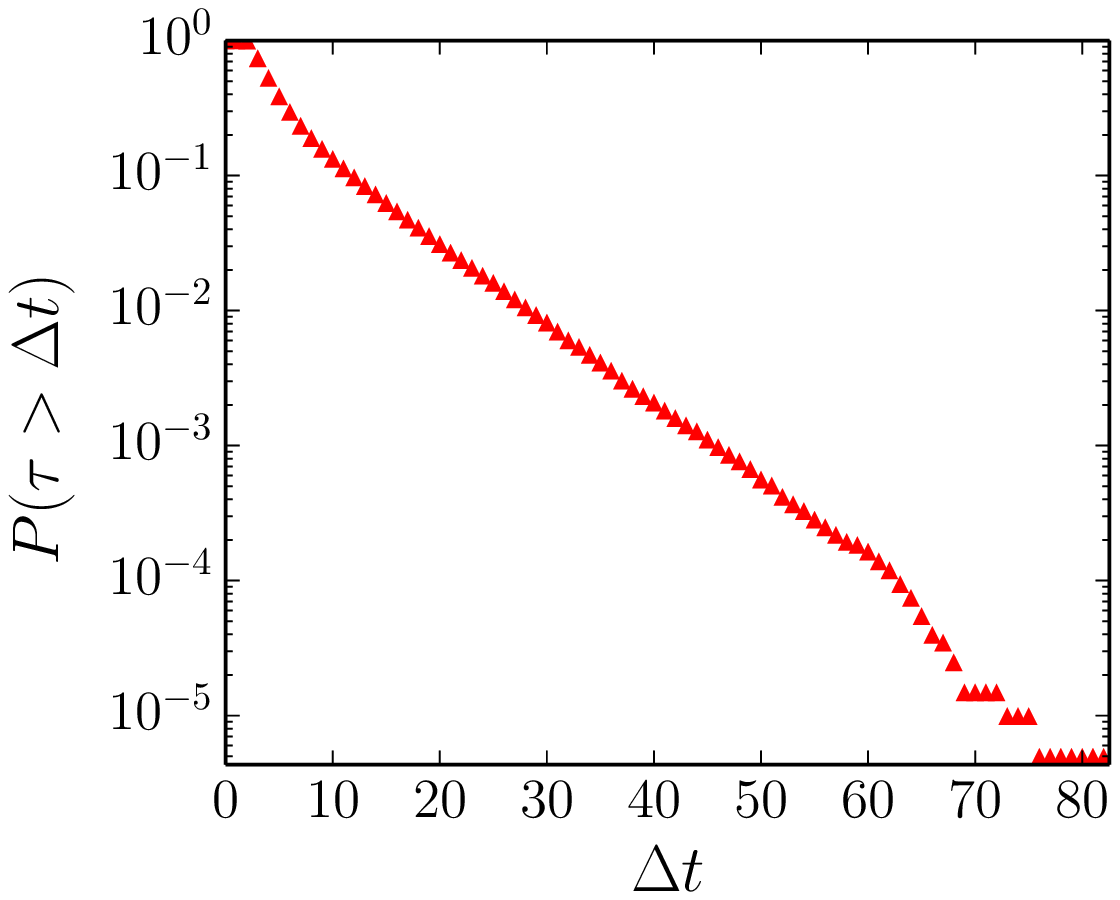}
\par\end{centering}

\caption{\label{fig:app_Fake_power_laws}Graphical representations of the
distribution
function obtained with the contagion-free model for a strongly competitive
crowd, where the heterogeneous distribution of $\Pi^\mathrm{intr}$ is replaced
by a Dirac peak. Left: probability distribution function in a logarithmic plot.
Right: survival function in a semi-logarithmic plot.}
\end{figure}

As stressed by Clauset et al. \cite{clauset2009power}, some care
needs to be taken when fitting empirical data with power laws, insofar
as simple graphical representations of the empirical distribution
function may be misleading. Indeed, before implementing the cellular
automaton described in the main text, we worked with a slightly different
version (which, in particular, did not feature disorder in the intrinsic
probabilities $\Pi^{\mathrm{(intr)}}$). A logarithmic plot of the
distribution of time lapses obtained with that model (see
Fig.~\ref{fig:app_Fake_power_laws}a)
made us think that the data were well described by a power-law tail.
However, after plotting the complementary cumulated distribution (i.e.,
the survival function), we realized that the data were in fact
better described by an exponential distribution (compare
Fig.~\ref{fig:app_Fake_power_laws}b with the red dotted lines in
Fig.~\ref{fig:dist_Deltat_no_contagion_Pi0} of the main text).

This experience cast some doubt in our minds about the observations
of power-law distributions of time lapses with cellular automata reported
in the literature \cite{song2005cellular,wei2006evacuation}. By plotting
the survival functions of the data shown in these works and applying
the methods of Ref.~\cite{alstott2014powerlaw,clauset2009power}
to compute the likelihood to be a power law, we found that the data
of Ref.~\cite{wei2006evacuation} and, to a lesser extent,
\cite{song2005cellular}
are at least as compatible with an exponential tail as they are with
a power-law one.

\end{document}